\documentclass[epjCONF]{svjour}
\usepackage{graphics}
\usepackage[varg]{txfonts} 
\usepackage[latin1]{inputenc}
\usepackage{lineno}
\session-title{International Symposium on Future Directions in UHECR Physics 2012}
\begin{document}

\title{Review of the Multimessenger Working Group at UHECR-2012}
\author{J.~Alvarez-Mu\~niz \inst{1}  
and M.~Risse \inst{2} for the Pierre Auger Collaboration \\
G.I.~Rubtsov \inst{3}  
and B.T.~Stokes \inst{4} for the Telescope Array Collaboration}


\institute{Depto. de F\'\i sica de Part\'\i culas $\&$ Instituto Galego de F\'\i sica de Altas Enerx\'\i as,
Universidade de Santiago de Compostela, 15782 Santiago de Compostela, Spain \and 
University of Siegen, Department of Physics, 57068 Siegen, Germany \and 
Institute for Nuclear Research of the Russian Academy of Sciences, Moscow 117312, Russia \and 
University of Utah, High Energy Astrophysics Institute, Salt Lake City, Utah, USA.
}

\abstract{
The current status of searches for ultra-high energy neutrinos and
photons using air showers is reviewed. Regarding both physics and
observational aspects, possible future research directions are indicated.
} 

\maketitle

\linenumbers
\section{Introduction}
\label{intro}

What is the status of searches for ultra-high energy neutrinos and
photons using air showers? What might be the future prospects, in
particular in the next couple of years? What is (are) the physics
case(s) for multimessenger observations, and what are the
observational experiences and challenges? These questions may
summarize the main objective of the Multimessenger Working Group
that was formed, together with four other Working Groups, a few
months before the UHECR-2012 symposium. At this symposium, possible
future directions of the field of ultra-high energy cosmic rays
were discussed, bringing the major collaborations from air-shower
experiments as well as colleagues from theory together.

Given this objective, one can think of many issues to inspect.
In particular, one can compare neutrinos versus photons;
neutrinos and photons versus charged cosmic rays;
air shower observations versus other techniques;
shower observations from ground versus those from space;
ground shower techniques versus each other;
current data versus models; various models versus each other;
and, above all, the present status versus future directions.
Given these many aspects in a highly dynamic field, it is evident
this review does not claim to be complete, or even finished.
Rather, certain considerations of possible relevance to the aim
of the symposium are compiled and highlighted.

We start with both neutrinos and photons
before each one is examined individually.

\section{UHE Neutrinos and photons}
\label{sec:2}

Multimessenger observations are a key ingredient for discovering and
for better understanding various phenomena in the Universe.
Boosted by the invention of the telescope about 400 years ago,
photon observations cover now an impressive energy range from radio
wavelengths up to about 100 TeV.
The discovery of (charged) cosmic rays dates back (at the time of
writing these proceedings quite precisely) 100 years ago, with cosmic
rays being measured now from sub-GeV to more than 100 EeV in energy.
First non-terrestrial neutrinos were observed about 40 years ago,
in the MeV energy range.

Various efforts are underway for opening new observational windows
to the Universe and for deeper observations of already accessible energy
regimes of the different particle types.
Examples of significant discoveries in new windows are the cosmic 
microwave background or gamma-ray bursts for the case of photon
messengers, the muon for the case of cosmic-ray messengers via the air
showers they produce, or the discovery of neutrinos from SN 1987A in the case of neutrino
messengers. Two things can be seen from this small list. Firstly, some of
the discoveries were not expected or predicted beforehand; they gave
their rewarding justification for the efforts made to proceed into unknown
regime only afterwards: an important feature that also decision-makers
should keep in mind. And secondly, these discoveries impacted (or even
helped to establish) different fields of nowadays' physics, including
astrophysics, particle physics, cosmology and fundamental physics.
In this sense, astroparticle physicists should remain open-minded as
to what kind of physics they can ``do'' by using the astroparticles.

In the remainder of this section, we sketch the ``life'' of UHE neutrinos
and photons from production over propagation to searching for them using
air showers.  Along this path, we will see structural similarities and
differences in their appearance as multimessenger particles.

\subsection{Production.}

Both UHE neutrinos and photons are thought to be typically produced as
decay secondaries, i.e.\ they come from higher-energy cosmic rays.
These cosmic rays (nucleon or nucleus) may interact with matter, e.g.\
with gas around a source, or with background photons, e.g.\ the CMB
or photon fields around a source. The produced secondaries include in
particular pions, and neutrinos (photons) can be generated in the decay
of the charged (neutral) pion.
An important example is the GZK process, where a proton above 50 EeV
interacts with the CMB and gives rise (in about 1/3 of the cases)
to finally 3 neutrinos of about $5\%$ each of the initial proton energy,
or (in about 2/3 of the cases) to 2 photons of about $10\%$ each of the initial
proton energy.

As alternative scenarios, top-down models were proposed where the
pions can emerge from the decay or annihilation of exotic particles;
as will be commented on below, these models are now strongly constrained
by searches for UHE photons and neutrinos.
In any case, both messengers can emerge from the same type of
initial process, and finally from (pion) decay.

\subsection{Propagation.}

Both UHE neutrinos and photons propagate along a straight line and are
not deflected by magnetic fields, in contrast to charged cosmic rays.
This opens the possibility of doing ``astronomy'' by directional pointing.
The neutrinos, to good approximation, also do not interact. For
propagation over cosmological distance, maximum mixing of flavors is
typically assumed. UHE photons, with an energy loss length of order
of 10 Mpc at around 10 EeV, initiate an electromagnetic cascade down to
GeV-TeV energy. Thus, there is complementarity due to the different
interaction cross sections between the messengers, with UHE photons
testing the more local Universe, while UHE neutrinos (as well as
down-cascaded GeV photons) can reach us
from cosmological distances (assuming standard physics in all cases).

\subsection{Searches using air showers.}

Both UHE neutrinos and photons can be searched for using air showers.
In the case of UHE photons, fairly ``normal'' showers (compared to
hadron showers) are produced and typical shower detectors can be used.
The separation of photon-induced and hadron-induced showers
is based on composition-sensitive shower observables. This might be
a larger challenge for currently planned observations from space
compared to ground-based experiments. Overall, there is no competitive
technique to shower detectors so far to search for UHE photons.

In the case of UHE neutrinos, the probability to initiate an air shower
at all is quite small ($\sim$10$^{-5}$ at 1~EeV within a depth of
1000~g~cm$^{-2}$). But if a shower is produced, also the extreme
phase space can be populated, e.g.\ near-horizontal showers starting
at very large depths, or upward-going showers starting in the
Earth's crust. Due to this, a strong background reduction is possible,
with the limitation of trying to keep a large exposure. This might be
an opportunity for future space observations to try to add significant
exposure. Overall, the search for UHE neutrinos using air showers is
in competition with other techniques, e.g.\ with neutrino telescopes
using ice or water as a medium.

For both UHE neutrinos and photons, no discovery has been claimed so
far, and upper limits were derived.
In case of UHE photons, candidate events exist that are
usually close to the detector threshold and compatible with
expectations from hadronic background. For conservative limits,
this background is typically not subtracted to avoid uncertainties
in modeling high-energy hadronic interactions.
In case of UHE neutrinos, no candidates emerged so far from
searches using air showers such that the current search is not limited by
background but by exposure. 

For deriving limits, in both cases a fairly robust interpretation
of shower observations with regard to theory uncertainties is possible,
which is in some contrast to the more uncertain interpretation of hadron
showers: photons showers are mostly just electromagnetic. And for
neutrinos, while there is an uncertainty in cross-section and $\tau$
energy loss, the uncertainty in shower development is less important
since the identification relies mostly on the electromagnetic
component of the shower.

\subsection{Previous impacts.}

\begin{figure}
\resizebox{0.85\columnwidth}{!}{\includegraphics{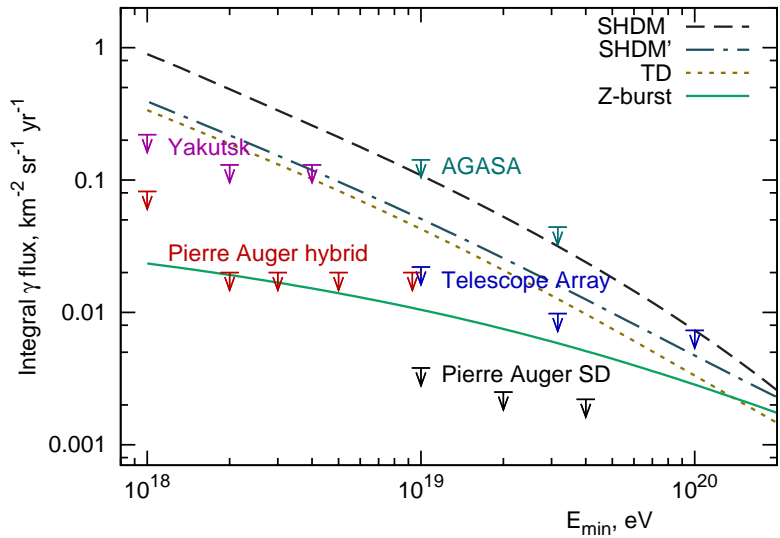}}
\caption{Constraints on selected top-down models of UHE photon 
production by current UHE photon flux limits.  
See Section \ref{sec:3} for more details on the photon limits.
 }
\label{fig:exotics_photons} 
\end{figure}

\begin{figure}
\begin{center}
\resizebox{0.75\columnwidth}{!}{\includegraphics{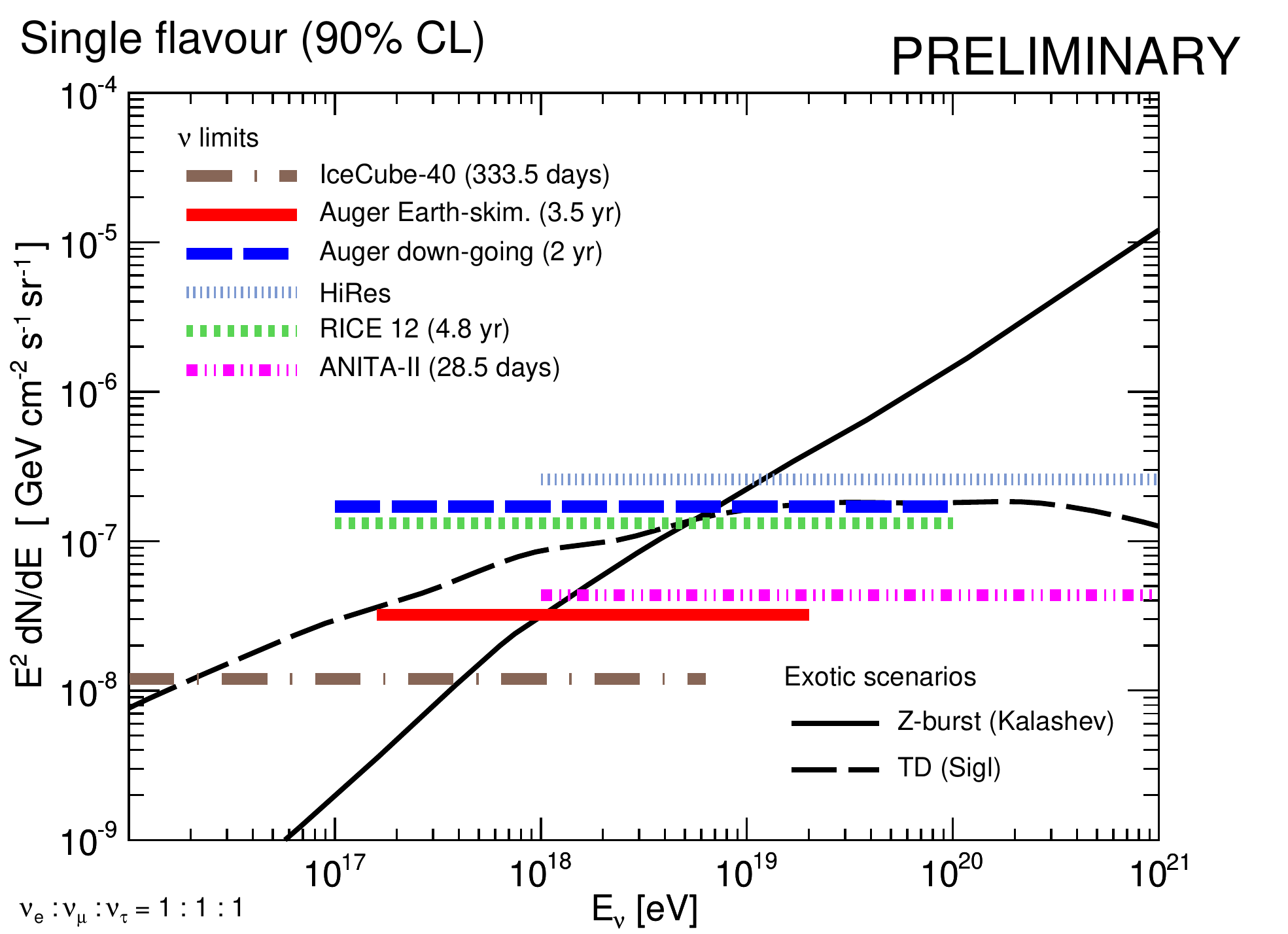}}
\caption{Constraints on selected top-down models of UHE neutrino
production by current UHE neutrino flux limits.  
See Section \ref{sec:4} for more details on the neutrino limits.
 }
\label{fig:exotics_nus} 
\end{center}
\end{figure}

Searches for UHE neutrinos and photons gave already
important contributions to the scientific landscape.
As can be seen in Figs.~\ref{fig:exotics_photons} and \ref{fig:exotics_nus}, 
strong constraints on top-down models
could be achieved by the photon, as well as the neutrino limits.
Further impacts 
will be discussed below. 

\section{UHE photons}
\label{sec:3}

Photon-induced showers are mostly electromagnetic with the first
interaction dominated by electron-positron pair production in the
Coulomb field. At the highest energies (above 10 EeV) two additional
effects come into play: the Landau, Pomeranchuk \cite{LP} and Migdal
\cite{M}~(LPM) effect and preshower in the geomagnetic field
(see~\cite{RisseRev} for a review). Photon-induced showers are deeper
and have significantly less muons compared to hadronic showers. The
observables used for photon-hadron separation include the muon density at
ground level, the depth of shower maximum ($X_{\rm max}$) and the
properties of the shower front (curvature, rise time,...), see
Table~\ref{tab:photons}. The best photon-hadron separation is achieved
using the muon detection technique, which at the same time has the
highest price per exposure. A hybrid technique using fluorescence and muons could 
possibly perform even better. All methods result in a merit factor
(defined as the ratio of separation power and price per exposure) of the same
order. 
Generally, any technique
good for hadronic composition study is adequate for photon search 
(as long as electromagnetic showers can trigger the detectors). 


\subsection{Present status}

\begin{figure}
\resizebox{0.85\columnwidth}{!}{\includegraphics{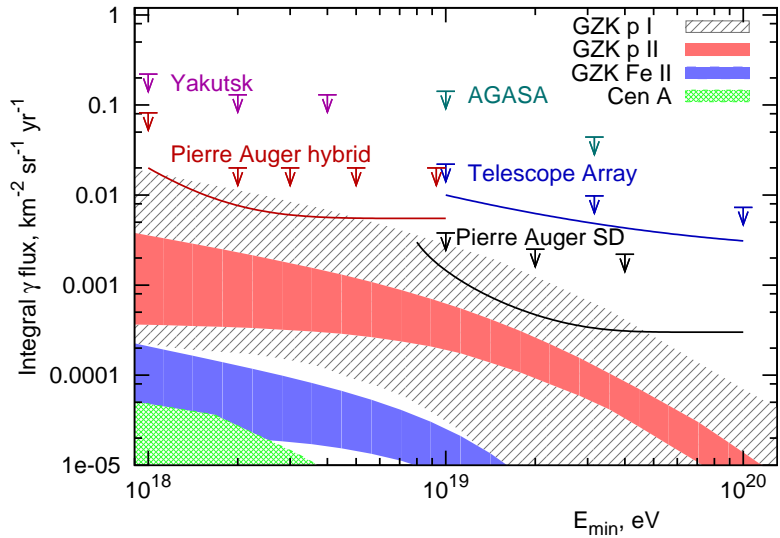}}
\caption{Limits to the UHE photon flux placed by several experiments 
  namely,  
  AGASA~\cite{AGASA_1stlim}, Yakutsk~\cite{Ylim18}, Pierre
  Auger~\cite{Auger_sdlim,Auger_hyblim2} and Telescope
  Array~\cite{TAglim} experiments. 
Also shown are estimates of the sensitivity with data until 2015 
as derived by scaling the current limits to account for the relative expected increase of the exposure,
and assuming the number of background events remains constant.
The flux predicted in several cosmogenic
  models of UHE photon production ~\cite{models,models2} and a Cen A source model~\cite{CenA} are shown
  in the shaded region.
 }
\label{fig:g2015} 
\end{figure}

The findings of independent experiments in both North and South
hemispheres are similar: no photon candidates were found at the highest
energies and the number of candidates found 
can be attributed to hadronic background (deep proton-induced showers). The existing
photon flux limits along with the predictions of the models are shown
in Fig.~\ref{fig:g2015}. 
Also shown are estimates of the sensitivity with data until 2015 
as derived by scaling the current limits to account for the relative expected increase of the exposure,
and assuming the number of background events remains constant.
The flux of GZK photons critically depends on
the source model, and the existing limits are getting close to the
predictions of the most optimistic scenarios (with proton primaries).

\begin{figure}
\resizebox{0.85\columnwidth}{!}{\includegraphics{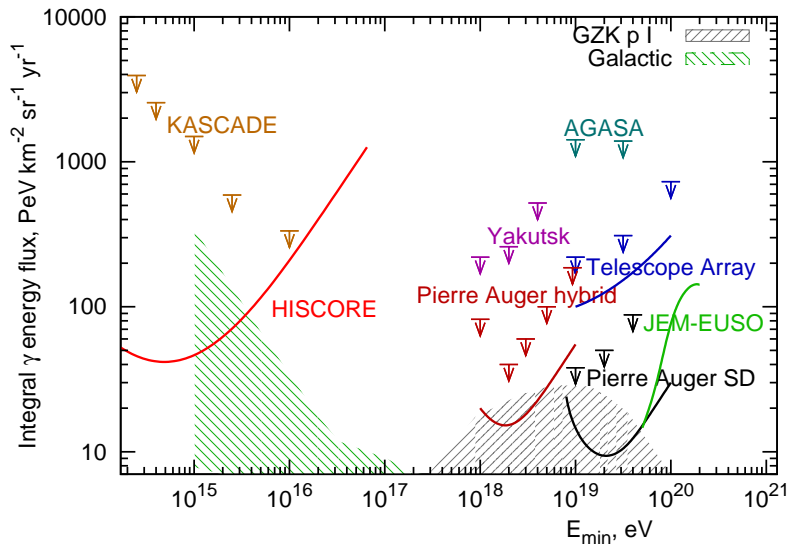}}
\caption{Limits to the photon flux from $\sim 10^{14}$ eV to $\sim 10^{20}$ eV,
  including the results of KASCADE~\cite{KASCADElim} and
  the expected sensitivities of the Pierre Auger Observatory and Telescope Array by 2015, HiSCORE~\cite{HiSCORE} and
  JEM-EUSO~\cite{JEMEUSO}. The prediction for the flux of galactic gamma rays 
  at PeV energies is taken from Ref.~\cite{Galactic}. }
\label{fig:gbroad} 
\end{figure}

Presently there is an unexplored energy gap $10^{16}$--$10^{18}$~eV
between the flux limits at lower energies established by
KASCADE~\cite{KASCADElim} and the limits set in UHECR experiments, see
Fig.~\ref{fig:gbroad}. The gap is a target for the low energy
extensions of Pierre Auger and Telescope Array observatories and
future large scale air Cherenkov detectors
(e.g. HiSCORE~\cite{HiSCORE}).

\subsection{Possible Impact of UHE Photon Searches}
\label{sec-impact}

Photons, as the gauge bosons of the electromagnetic force, at such
enormous energy can be regarded as unique messengers and probes of
extreme and, possibly, new physics.  Implications are related to the
production of photons, their propagation, and interactions at the
Earth.  Many aspects of the following, incomplete list (cf. \cite{RisseRev}) of possible
impacts of UHE photon searches and connections to other research
subjects require more study.

Large UHE photon fluxes are a {\it smoking gun} for current
non-acceleration models.  Stringent photon limits give parameter
constraints such as a lower limit on the lifetime of relic SHDM
particles \cite{aloisio,Kalashev:2008dh}.

Findings on photons are needed to reduce corresponding systematics in
other air shower studies, such as the primary composition, energy
spectrum or when trying to constrain interaction parameters such as
the proton-air cross-section~\cite{belov,ulrich} from showers.

UHE photons may be helpful for diagnostics of sources accelerating
nuclear primaries, as the photon fluxes from UHE hadron interactions
are expected to be connected with source features such as the type of
primary, injection spectrum, possible beam dump at the source, or
source distribution (see also \cite{models,models2}).

UHE photons point back to the location of their production.  Possibly,
the arrival directions of photons may correlate better with the source
direction than those of charged primaries.  There may be an enhanced
UHE photon flux from the galactic center region depending on the
spectra of nuclear primaries \cite{stecker06}. In certain SHDM
scenarios, an enhanced flux of $\sim$$10^{18}$~eV photons from the
galactic center is possible without a higher-energy
counterpart~\cite{berez_priv}.

Propagation features of UHE photons are sensitive to the MHz radio
background~\cite{sarkar03}. The photon flux at Earth is also sensitive
to extragalactic magnetic fields~\cite{sigl95}.

Already a small sample of photon-induced showers may provide
relatively clean probes of aspects of QED and QCD at ultra-high energy
via the preshower process and photonuclear interactions (see, e.g., \cite{rissec2cr}).

There are several connections to Lorentz invariance
violation \cite{Coleman:1998ti,liv1,liv2,liv3}. The production of GZK photons can be
affected as well as interactions of photons during propagation and
when initiating a cascade at the Earth.  Particularly, photon
conversion (interaction with background fields or preshower process)
may be suppressed. The observation of GZK secondaries may set the
strongest limits on the Lorentz invariance violation at Planck
scale~\cite{Galaverni:2007tq}.

It is interesting to check whether UHE photon propagation could be
affected by the presence of axions or scalar bosons. The formal
requirements for photon-axion conversion regarding photon energy and
magnetic field strength, appear to be
fulfilled~\cite{gabrielli1,gabrielli2}. Photon conversions to
non-electromagnetic channels may differ from the standard QED process
due to an absence of electromagnetic sub-cascade. Photon-axion
conversion may ultimately increase the propagation distance of the UHE
photons thus allowing to identify distant
sources~\cite{Fairbairn:2009zi}.

UHE photon propagation can be modified in certain models of brane
worlds~\cite{branes} quantum gravity theory~\cite{quantum1,quantum2}
or spacetime foam \cite{klinkhamer1,Klinkhamer:2010pq}.  For instance,
depending on fundamental length scales significant scattering of
photons on structures (defects) in spacetime foam can occur.  In turn,
constraints may be derived when actually observing UHE photons, even
with one gold-plated event only \cite{klinkhamer1,Klinkhamer:2010pq}.

\begin{table}
\caption{UHE photon flux limits and corresponding techniques.}
\label{tab:photons}
\begin{tabular}{llll}
\hline\noalign{\smallskip}
Experiment & Technique & Observables & Ref \\
\noalign{\smallskip}\hline\noalign{\smallskip}
Haverah Park & SD: water Cherenkov & attenuation of inclined showers & \cite{HP_lim} \\
AGASA & SD: scintillator \& muon & muon density & \cite{AGASA_1stlim,AGASA_Risse}\\
Yakutsk & SD: scintillator \& muon & muon density & ~\cite{Ylim,Ylim18,A+Y}\\
Pierre Auger  & SD: water Cherenkov & front curvature, rise time & ~\cite{Auger_sdlim}\\
Observatory   & hybrid: fluorescence + SD & $X_{max}$, particle density far from the core & ~\cite{Auger_fdlim,Auger_hyblim2}\\
Telescope Array & SD: scintillator & front curvature & ~\cite{TAglim}\\
\noalign{\smallskip}\hline
\end{tabular}
\end{table}

\section{UHE neutrinos}
\label{sec:4}

The observation of UHE neutrinos (UHE$\nu$s) in the EeV energy range and above
has become a priority in experimental Astroparticle Physics. The recent observation 
of two candidates in the 1-10 PeV energy range with the completed IceCube detector \cite{IceCube_Nu2012} 
- still of unknown origin - 
encourages the search for these elusive particles with ground arrays of particle detectors. 

\subsection{UHE neutrino detection}

The observation of UHE$\nu$ in ground arrays is currently limited by exposure but not 
by background. UHE$\nu$s can induce extensive air showers that   
populate regions of phase space in zenith angle and injection depth in the atmosphere that 
are very unlikely to be accessed 
by showers initiated by UHECRs and photons. Using Monte Carlo simulations it has been established that neutrino identification 
at ground arrays can be performed with a large efficiency as long as the search is restricted to very inclined 
showers (typically above $60^\circ$ zenith angle), starting deep in the atmosphere close to ground \cite{Berezinsky_HAS}, 
and to upward-going showers \cite{nutau}. 
Since the injection depth is not directly measured in ground arrays, other surrogate observables are used.  
Unlike UHECR-induced cascades, a nearly-horizontal neutrino-induced shower initiated close to ground will 
have a significant electromagnetic component at the detector. As a consequence the shower front is typically 
broader in time than the corresponding one in a UHECR-induced shower interacting high in the atmosphere 
which is mainly constituted by muons. 
This is the basis for the identification of neutrino candidates in all ground arrays currently 
operating in the UHE range, namely the Pierre Auger Observatory 
and the Telescope Array, and it calls for detectors with both sensitivity to highly inclined showers 
and timing capabilities. 

The sensitivity of ground arrays extends to all neutrino flavors and all type of 
interactions (charged-current CC or neutral-current NC) relevant at UHE energies \cite{nutau_zas}. 
Neutrinos of electronic, muonic and tauonic flavor can collide with nuclei in 
the atmosphere and induce an EAS close to the ground that can be identified in 
a broad zenith angle range from $\theta \sim 60^\circ - 90^\circ$. In this so-called 
``downward-going" neutrino channel, all flavours and both CC and NC interactions
contribute to the neutrino event rate. The sensitivity to UHE tau neutrinos is 
further enhanced through the detection of the shower induced by the tau lepton 
generated after the propagation and interaction of an upward-going $\nu_\tau$ 
inside the Earth \cite{nutau}, the so-called Earth-skimming mechanism. 
In this case only CC $\nu_\tau$ interactions can be efficiently detected. 
The angular range in which this technique is viable at EeV energies is restricted 
to $\theta= 90^\circ - \sim 95^\circ$. At larger zenith angles the Earth is opaque 
to UHE $\nu_\tau$ and/or the shower emerging from the Earth is too upward-going to reach ground.  
Despite these limitations, the Earth-skimming mechanism is roughly a factor $2.5-3$ more 
sensitive than the downward-going one in the EeV energy range, mainly due to the $\sim$ 2000 times larger
density of the target for neutrino interactions (the Earth crust) when compared to the atmosphere.
The possibility to detect Earth-skimming neutrinos of electronic flavor has also been explored,
but the sensitivity is expected to be smaller than other channels \cite{HiReS_nue,Tartare}.
This information is summarized in Table \ref{tab:nu-channels}.

\begin{table}
\caption{
Channels, $\nu$ flavors and interactions that can be detected in ground arrays of particle detectors 
in the EeV range. Also shown are the zenith angle ranges where the two main channels (Earth-skimming
and downward-going) are most sensitive, as well as their relative sensitivity.}
\label{tab:nu-channels}
\begin{tabular}{lll}
\hline\noalign{\smallskip}
Channel                     & Earth-skimming                 & Downward-going                             \\  
\noalign{\smallskip}\hline\noalign{\smallskip}
Flavours \& interactions    & $\nu_\tau$ CC                  & $\nu_e$, $\nu_\mu$, $\nu_\tau$ \& CC + NC  \\ 
Zenith angle range          & $90^\circ - \sim 95^\circ$     & $\sim 60^\circ - 90^\circ$                 \\ 
Target for $\nu$s           & Earth's crust                  & Atmosphere                                 \\ 
Density [${\rm g~cm^{-3}}$] & $\rho\sim 2-3~{\rm g~cm^{-3}}$ & $\rho\sim 10^{-3}~{\rm g~cm^{-3}}$         \\ 
Relative sensitivity        & $\sim 2.5 - 3$                 & 1                                          \\ 
\noalign{\smallskip}\hline
\end{tabular}
\end{table}

In principle the search for UHE neutrinos can also be done with fluorescence detectors 
which are directly sensitive to the shower longitudinal profile 
and can in principle identify very penetrating inclined showers. 
However fluorescence telescopes can only work during moonless
nights, and this limits their duty cycle to approximately 10-15$~\%$ of the total time. 
As a consequence, this reduces the exposure to UHE$\nu$s compared to that of ground arrays
of particle detectors. 

Other techniques for UHE neutrino detection are being applied in other experiments \cite{Veronique}. 
Neutrinos can interact in a ``dense" medium such as water or ice, and the charged debris 
of their collisions - that typically travel at a speed larger than the speed 
of light in the media - emit Cherenkov light. 
This technique is exploited in experiments such as IceCube and ANTARES, 
which benefit from the abundance and relatively large density of the target,
and from the transparency of water and ice to visible wavelengths which allows a sparse array of 
detectors buried/submerged inside the target itself.  
Cherenkov radiation is also emitted in the MHz-GHz frequency range. 
These wavelengths are typically larger than the dimensions of the neutrino-induced shower 
in a dense medium, and radiation is emitted coherently, with the power in radiowaves scaling
as the square of the $\nu$ energy. This so-called radio technique
has been already exploited in experiments which attain their maximum sensitivity in the UHE regime,
such as ANITA and RICE, and it is starting to be explored in initiatives such as ARA \cite{ARA} and ARIANNA \cite{ARIANNA}.
The radio technique is also the basis of neutrino detection in experiments using existing radio telescopes
to try to detect radio pulses produced in neutrino-induced cascades in the Moon.  
In Table \ref{tab:limits} we summarize the main characteristics of the 
different UHE neutrino detection techniques.

\begin{table}
\caption{Description of $\nu$ detection techniques in the EeV range along with limits to the diffuse flux
of UHE$\nu$s (when available in the bibliography).}
\label{tab:limits}
\begin{tabular}{lllll}
\hline\noalign{\smallskip}
           & Technique                     &                   & Single flavour limit  (90$\%$ C.L.) &    \\ 
Experiment & $\nu-$flavors                 & Observables       & $[{\rm GeV~cm^{-2}~s^{-1}~sr^{-1}}]$  & Ref.\\ 
           & Target                        &                   & Livetime / $E_\nu$-range (EeV)              &     \\ 
\noalign{\smallskip}\hline\noalign{\smallskip}
Pierre       & Array of water stations     & Shower zenith angle     &                                             &         \\  
Auger        & Earth-skimming $\nu_\tau$   & Time-structure of front &  {$k= 3.2~10^{-8}$}         & \cite{Auger_ES} \\  
Observatory  & Earth's crust               &                         &   3.5 yr full Auger / $\sim$ 0.16 -- 20  &      \\ 
\hline\noalign{\smallskip}
Pierre       & Array of water stations     & Shower zenith angle        &                                        &         \\   
Auger        & Downward-going $\nu_{e,\mu,\tau}$   & Time-structure of front &  {$k= 1.7~10^{-7}$}    & \cite{Auger_DG}  \\  
Observatory  & Atmosphere, mountains       &                         &   2.0 yr full Auger / $\sim$ 0.1 -- 100  &        \\
\hline\noalign{\smallskip}
             & Fluorescence telescopes     & Shower incidence        &                   &         \\  
HiReS        & Earth-skimming $\nu_{\tau,e}$  & (upward-going) &  {$k \sim 2.6~10^{-7}$} & \cite{HiReS_nu_tau} \\ 
             & Earth's crust               &                         &  $\sim~$6.5 yr data taking / $\sim$ 1 -- 100 &         \\
\hline\noalign{\smallskip}
Telescope    & Array of scintillators      & Shower incidence        &                   &         \\ 
Array        & Downward-going $\nu_{e,\mu,\tau}$   & Time-structure of front & {--}         & \cite{TA_ICRC11} \\ 
             & Atmosphere, mountains       &                         & $\sim~$3 yr data taking / $>$ 1  &         \\ 
\hline\noalign{\smallskip}
             & 3D array buried PMTs        & Shower or $\mu$ incidence &                             &         \\  
IceCube      & Up/Horizontal $\nu_{e,\mu,\tau}$      & Cherenkov light    & {$k= 1.2~10^{-8}$} &  \cite{IceCube} \\ 
             & Ice                         &                           &  333.5 days / $2.0~10^{-3}$ -- 6.3     &         \\
\hline\noalign{\smallskip}
             & Antenna Balloon             & Shower incidence          &                  &         \\ 
ANITA-II     & Upward-going $\nu_{e,\mu,\tau}$         & Cherenkov at MHz -- GHz   & {$k= 4.3~10^{-8}$} & \cite{ANITA} \\ 
             & Antarctic Ice               & (Askaryan radiation)      &  28.5 days / 1.0 -- $3.1~10^5$      &         \\ 
\hline\noalign{\smallskip}
             & Array buried antennas       & Shower incidence          &                  &         \\  
RICE         & Upward-going $\nu_{e,\mu,\tau}$         & Cherenkov at MHz --  GHz  & {$k= 1.3~10^{-7}$} & \cite{RICE} \\ 
             & Antarctic Ice               & (Askaryan radiation)      &  4.8  yr / $\sim$ 0.1 -- 100        &         \\
\hline\noalign{\smallskip}
RESUN        & Radio telescope             &                           & $k \sim 7.5~10^{-6}$          & \cite{RESUN} \\
             &                             & Cherenkov at MHz-GHz      &  200 h. / $\sim~1.6~10^3$ -- $3.1~10^4$  &              \\ \cline{1-1} \cline{4-5} 
NuMoon       & Moon-skimming $\nu_{e,\mu,\tau}$         & from the Moon             &      --                                  & \cite{NuMoon} \\
             &                             &                           &  50 h./ $>~4~10^4$                     &               \\ \cline{1-1} \cline{4-5} 
LUNASKA      & Moon regolith               &                           &  $k \sim 1.0~10^{-5}$                     & \cite{LUNASKA,Clancy} \\  
             &                             &                           &  33.5 h./  $\sim~10^3$ -- $10^5$                          &                \\  
\noalign{\smallskip}\hline

\end{tabular}
\end{table}

\subsection{UHE neutrino production models}

\begin{figure}
\begin{center}
\resizebox{0.75\columnwidth}{!}{\includegraphics{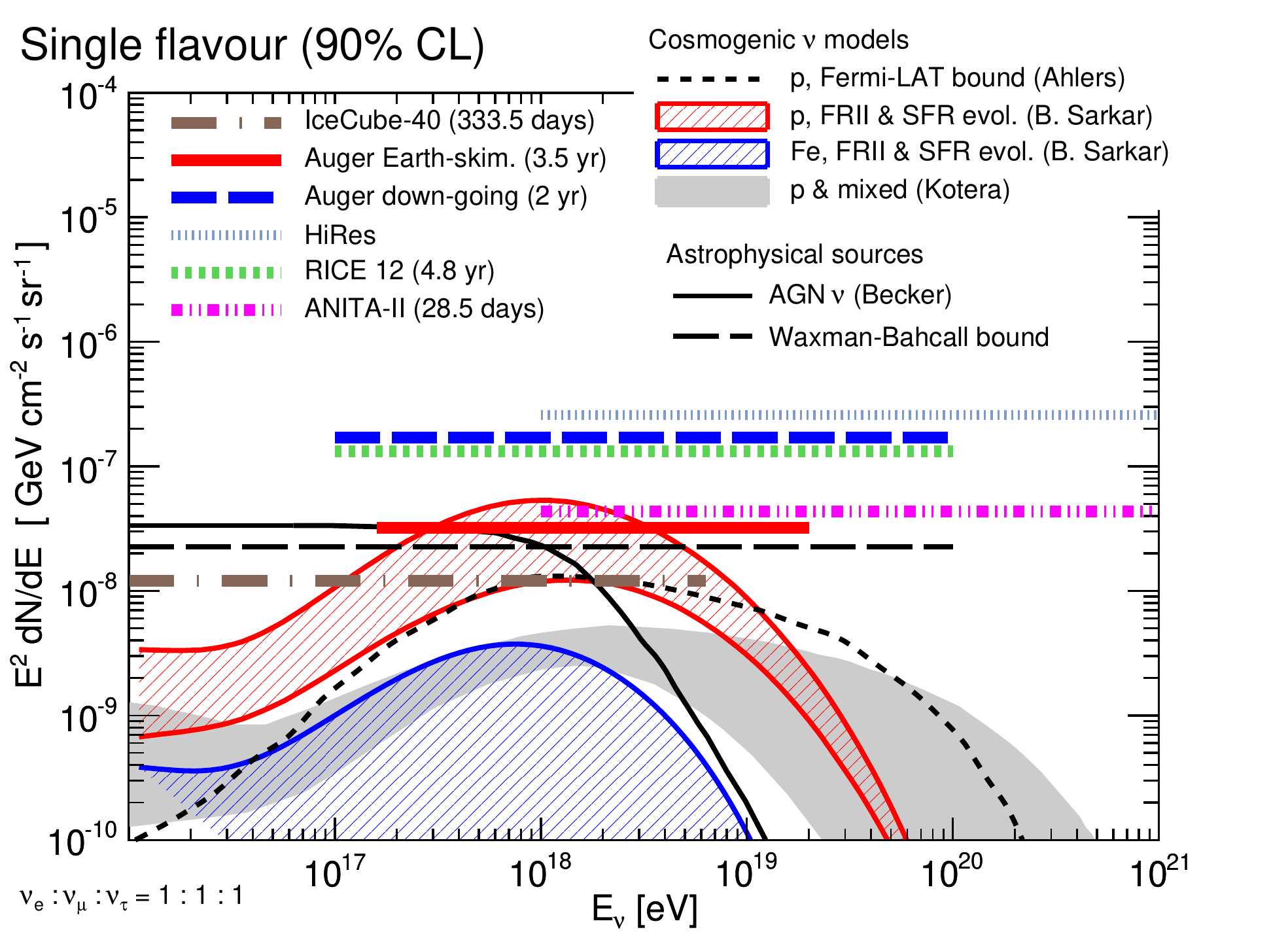}}
\caption{Expected fluxes for several models of cosmogenic neutrinos \cite{Kotera_GZK,Ahlers_GZK,Kampert_GZK}. 
The upper (lower) edge of the red band at the top corresponds to a model where proton primaries are 
injected at the sources which are assumed to follow a strong-FRII (weak-Star Formation Rate) evolution with redshift. 
The blue band at the bottom assumes the same but for iron primaries at the sources. In both cases power law source distributions with an 
injection index of $\gamma = -2$ have been assumed and a maximum energy of $E_{\rm max}= Z \times 10^{20}$ eV \cite{Kampert_GZK}.
The gray band represents a set of models with pure proton and mixed compositions at the sources,
and different assumptions on the evolution of the sources as well as on the transition from 
Galactic to extragalactic sources \cite{Kotera_GZK}. The dashed line is the cosmogenic $\nu$ model 
in \cite{Ahlers_GZK} (best-fit and $E_{\rm min}=10^{19}$ eV) constrained by Fermi-LAT 
observations of the GeV-TeV diffuse $\gamma$-ray background. 
The solid black line corresponds to a model of neutrino production in AGN \cite{BBR}. The dashed horizontal line 
is the Waxman-Bahcall bound for redshift evolution of the sources. 
Also shown are the integrated upper limits (at $90\%$ C.L.)
to the diffuse flux of UHE neutrinos (assumed to behave with energy as $dN/dE=k~E^{-2}$) 
from experiments with sensitivity to $\nu$s in the EeV range and above. 
See Table~\ref{tab:limits} for full details and relevant references. 
The Auger downward-going limit was obtained with data between 
zenith angles 75$^{\circ}$ and 90$^{\circ}$.
All limits and flux models have been rescaled to single flavour when necessary, 
assuming equipartition of flavors at Earth.}
\label{fig:1_nu}
\end{center}
\end{figure}

All experiments working in the EeV range aim at detecting the cosmogenic neutrino flux 
produced in interactions of UHECRs above $\sim$ 50 EeV with the Cosmic Microwave Background (CMB) \cite{BZ}. 
Despite the existence of these neutrinos should be guaranteed by the observation of both projectiles of 
that energy and target photons, the level at which this flux is realized in Nature is dependent 
on many unknown parameters such as the composition of UHECR at the sources, the spatial distribution of the 
sources and their evolution with time, the features of the UHECR spectra at the production sites, etc \cite{Allard_GZK,Kotera_GZK}. 
The large phase space leads to a wide range of predictions as shown in Fig.~\ref{fig:1_nu}.
In particular if the primary UHECR flux at the sources is dominated by iron, the fluxes are at least
one order of magnitude smaller then in the case of proton dominated fluxes.
However cosmogenic neutrinos are key to constrain this parameter space. Observations of UHECR alone 
do not uniquely determine both the injection spectrum and the evolution of the sources, mainly because 
interactions of UHECR during propagation obscure the early Universe from direct observation. It is these
same interactions that produce the cosmogenic neutrinos that keep memory of the parameters of the sources \cite{Stanev_GZK}. 

The cosmogenic neutrino flux is accompanied by
electrons, positrons and gamma-rays that quickly cascade on the CMB and intergalactic magnetic
fields to lower energies and generate a $\gamma$-ray background in the GeV-TeV region.
By measuring this $\gamma-$ray flux the neutrino production rate can be constrained \cite{Berezinsky_GZK_Fermi}.
An example, shown in Fig.~\ref{fig:1_nu}, is the cosmogenic neutrino flux in \cite{Ahlers_GZK} 
which is constrained by the diffuse GeV-TeV $\gamma$-ray flux observed by the Fermi satellite \cite{Berezinsky_GZK_Fermi}.
UHE neutrinos can also be produced at the potential sources of UHECRs themselves, 
and in particular in Active Galactic Nuclei (AGN) \cite{Becker}. An example \cite{BBR} is shown in Fig.~\ref{fig:1_nu}.
A benchmark model included in Fig.~\ref{fig:1_nu} is the Waxman-Bahcall neutrino flux bound \cite{WB}, a theoretical 
bound obtained normalizing the energy density in neutrinos to the energy production rate in protons needed 
to sustain the observed UHECR spectrum above $10^{19}$ eV. 

\subsection{Present status}

The two main ground arrays of particle detectors currently working in the EeV range, namely the Pierre Auger Observatory in Argentina 
and the Telescope Array in Utah have reported no neutrino candidates in their data. 
This allows to put a limit to the UHE neutrino flux. Conventionally, limits are displayed 
in integrated and differential formats. In the integrated format an energy dependence 
of the neutrino flux following a power-law function $k\cdot E^{-2}$ is assumed. This flux is integrated
in energy with the energy-dependent exposure of the detector obtained through Monte Carlo
simulations, and the total event rate is calculated. A limit can then be put to the value of the flux magnitude 
$k$ that would be needed to produce $2.44$ events in the observation period, corresponding to the upper bound of the 
confidence belt at $90\%$ C.L. in the 
Feldman-Cousins treatment of a null detection \cite{Feldman-Cousins}. The integrated limits for different neutrino experiments
obtained in a similar way as explained here are shown as straight lines in Fig.~\ref{fig:1_nu}. 
\footnote{A candidate event has been reported by the ANITA Collaboration \cite{ANITA}. The event 
is fully compatible with background and is included in the final value of the integrated limit through the Feldman-Cousins 
confidence belt approach. Note also that for the IceCube case we show the limit obtained with the IceCube-40 configuration
with no candidates reported and published in \cite{IceCube}.}
Integrated limits to the flux of point-like potential sources of UHE$\nu$s also exist \cite{Auger_ES,IceCube_point}

\begin{figure}
\begin{center}
\resizebox{0.75\columnwidth}{!}{\includegraphics{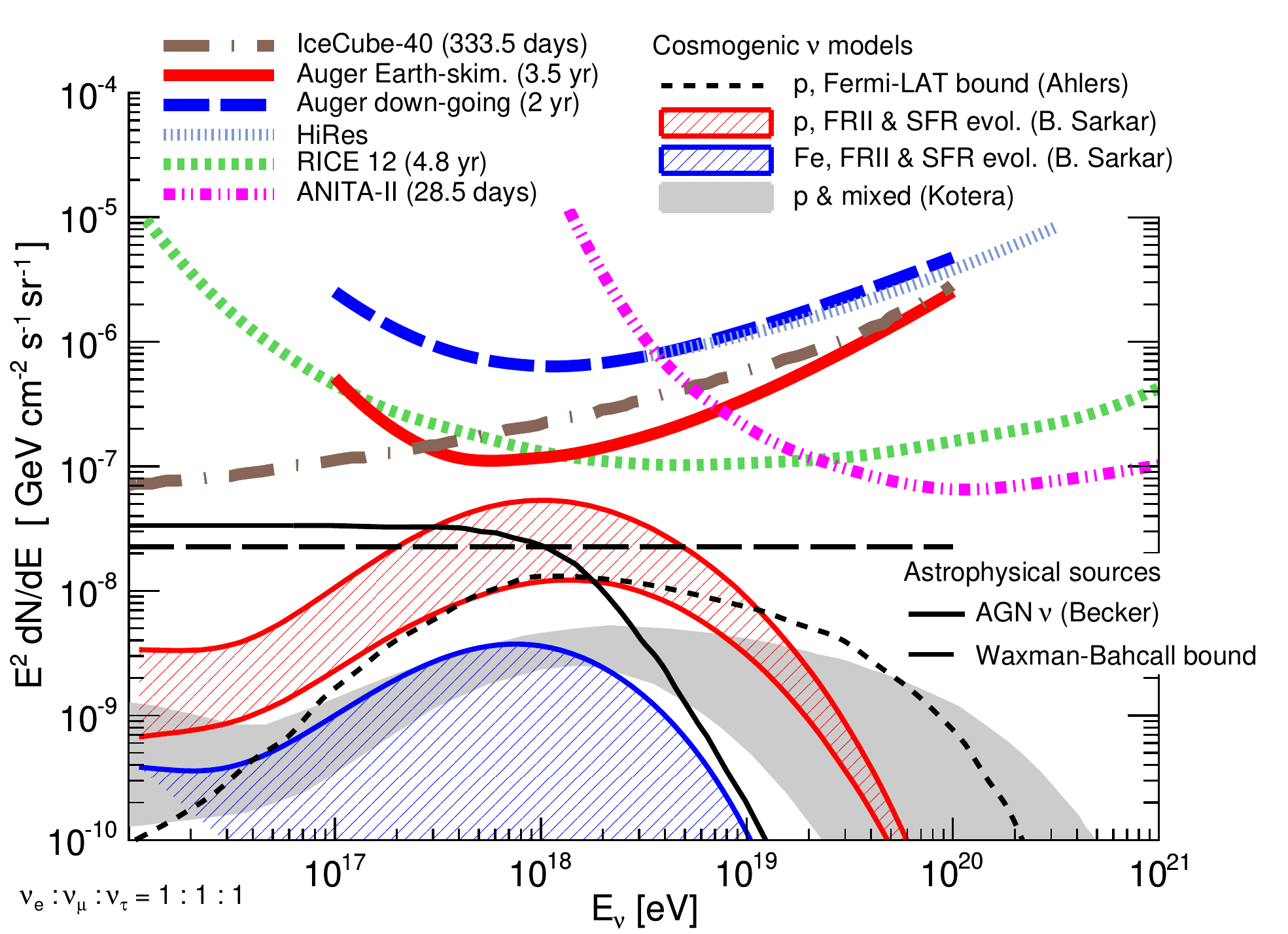}}
\caption{
Same as in Fig.~\ref{fig:1_nu} but with the limits  displayed
in differential format (see text for details). 
The limits have been scaled to single flavour
when necessary, and the IceCube differential limit
is further scaled down by a factor $1/2$ due to the different binning in energy with respect 
to the Auger differential limits. For the calculation of the RICE and ANITA differential limits  
we refer the reader to the original references \cite{RICE,ANITA}.
\label{fig:2_nu}
}
\end{center}
\end{figure}

The limits can also be shown in differential format. In this case the integrated limits are obtained 
in several energy bins of fixed width. 
These limits obtained by several experiments are shown in Fig.~\ref{fig:2_nu}.
This format has the advantage that it displays the energy range at which the experiments 
are most sensitive. As can be seen in Fig~\ref{fig:2_nu} the best sensitivity at the Auger Observatory 
is reached in the energy bin around 1 EeV which corresponds to the 
peak of the cosmogenic neutrino flux in an $E^2$ times flux plot, while for IceCube the 
largest sensitivity is achieved typically at 2-3 orders of magnitude smaller energy. On the contrary,
the ANITA sensitivity peaks between 1-2 orders of magnitude larger energy than Auger  
\footnote{The differential ANITA limit does not account for the background event.}. 

The best way to compare the sensitivity of the experiments is to quote the 
expected number of events during the data taking period of the detector for a few reference
models. This is shown in Table~\ref{tab:rates} for two representative cosmogenic 
neutrino flux predictions and the AGN neutrino model shown in Fig.~\ref{fig:1_nu}
and for the current exposures of IceCube-40 and the surface detector of the Auger 
Observatory (see Table~\ref{tab:limits} for more details on the exposures).
With data unblinded up to 31 May 2010, the surface detector of the Auger Observatory
provides a poor constraint to models of cosmogenic $\nu$ production normalized to Fermi-LAT observations.

\subsection{Future prospects}

As reflected in the projected number of events shown in Table~\ref{tab:rates}, 
if UHE$\nu$ are not discovered by $\sim$ 2015, experiments will be able to put
strong constrains on models of cosmogenic $\nu$ fluxes that assume a pure primary proton
composition injected at the sources.

\begin{table}
\caption{ 
Top: Expected number of $\nu$ events for several models of $\nu$ production given the current 
exposures of IceCube-40 and the surface detector of the Pierre Auger Observatory
to Earth-skimming $\nu$s only (see Table \ref{tab:limits} for details).
Bottom: Expected number of events for 3 yrs of data taking with the full IceCube detector, 
and for the Earth-skimming and downward-going channels at the SD of the Auger Observatory
up to June 2015.}
\label{tab:rates}
\begin{tabular}{lll}
\noalign{\smallskip}\hline\noalign{\smallskip}
Model / Detector                                & IceCube-40             & Auger (Earth-skimming)   \\ 
\hline\noalign{\smallskip}
Cosmogenic $\nu$ \cite{Ahlers_GZK}              &                        &                          \\
Primary p, Fermi-LAT constrained                & $\sim$ 0.4             & $\sim$ 0.6               \\ 
(dashed black line Fig.~\ref{fig:1_nu},\ref{fig:2_nu})         &                        &                          \\ 
\hline\noalign{\smallskip}
Cosmogenic $\nu$ \cite{Kampert_GZK}             &                        &                          \\
Prmary p, FRII source evolution                 & $\sim$ 1.8             & $\sim$ 2.2               \\
(top band top edge Figs.~\ref{fig:1_nu},\ref{fig:2_nu})         &                        &                          \\ 
\hline\noalign{\smallskip}
Active Galactic Nuclei $\nu$ \cite{BBR}         &                        &                          \\ 
                                                & $\sim$ 5.5             & $\sim$ 1.2               \\
(solid black line Figs.~\ref{fig:1_nu},\ref{fig:2_nu})          &                        &                          \\ 
%
\noalign{\smallskip}\hline\noalign{\smallskip}
Model / Detector                                & IceCube-86 (3 yrs)     & Auger 2015. Up + Down ($60^\circ-90^\circ$)  \\  
\hline\noalign{\smallskip}
Cosmogenic $\nu$ \cite{Ahlers_GZK}              &                        &                                \\
Primary p, Fermi-LAT constrained                & $\sim$ 2.3             & $\sim$ 2.5                     \\
(dashed black line in Figs.~\ref{fig:1_nu},\ref{fig:2_nu})      &                        &                                \\ 
\hline\noalign{\smallskip}
Cosmogenic $\nu$ \cite{Kampert_GZK}             &                        &                                \\
Primary p, FRII evolution                       & $\sim$ 10.3            & $\sim$ 9.2                     \\ 
(top band top edge Figs.~\ref{fig:1_nu},\ref{fig:2_nu})         &                        &                                \\ 
\noalign{\smallskip}\hline
\end{tabular}
\end{table}

\section{Conclusions}
\label{sec:5}

We have reviewed the status of searches for ultra-high energy
neutrinos and photons, touching different aspects:

\begin{itemize}

\item
Neutrinos versus photons:
Both are produced typically from the same initial process,
namely as secondaries from pion decay. Photons test
the local, neutrinos the whole Universe. Photons produce
fairly normal air showers (but with larger $X_{\rm max}$ and fewer
muons compared to hadronic ones), with small background
(deep proton showers). Neutrinos have a minute probability
(order $10^{-5}$) to produce an air shower at all, but if
so, it can be an extreme one (very deep or even upward)
with no background. The effective aperture ratio of a
ground array at 10~EeV is about $10^4$ comparing photons
to neutrinos.

\item
Neutrinos and photons versus charged cosmic rays:
Both neutrinos and photons are neutral which may allow
source pointing (``astronomy"). Both may act as messengers
of the GZK process. For both, the interpretation of
air shower observations is more robust than that of
hadron showers.

\item
Current data versus models: searches using air showers were already
performed and have already been useful by providing constraints on top-down models, on
Lorentz violation, and (starting) on astrophysics.
Discoveries of neutrinos and photons are well possible
though not guaranteed in the near future.

\item
Various models versus each other:
There is a large uncertainty concerning predictions.
It is and will be an on-going task for phenomenology
to combine the various constraints.

\item
Air shower observations versus other techniques:
For photon searches, there is no competing technique.
For neutrinos, air shower searches are (presently)
comparable to other techniques; possible findings by
other techniques may strongly impact future plans also
for neutrino searches by air showers.

\item
Shower observations from ground versus those from space:
The separation of primaries seems better from ground.
Some balance might come if large exposures can be reached
from space.

\item
Ground shower techniques versus each other:
For photon searches, each technique is OK with some
(dis-)advantages. As they have some complementarity,
the best would be a combination. In a simplified way, one can state
that what is good to determine (hadron) composition is
fine to search for photons. For neutrino searches,
a large exposure to inclined showers is the key, which
favours, e.g., a large array of water detectors.

\item
Present status versus future directions:
New observational windows to the Universe always gave new discoveries with large impact, also beyond
astroparticle physics. This is expected to continue
also for the windows which remain to be opened.
At the highest-energy frontier, the air shower community
has the means in hand to proceed towards the observation
of neutrinos and photons.

\end{itemize}



\end{document}